\documentstyle[12pt]{article}
\newcommand{\la}{\langle}
\newcommand{\ra}{\rangle}
\begin{document}
\begin{titlepage}
\begin{flushright}
FERMILAB-Pub-97/056-T\\
CERN-TH/97-33\\
\end{flushright}
\vspace{1cm}
\begin{center}
{\Large\bf On the Construction of Scattering Amplitudes for 
           Spinning Massless Particles }\\
\vspace{1cm}
{\large
F.~A.~Berends}\\
\vspace{0.5cm}
{\it
CERN TH-Division, CH 1211 Geneva 23, Switzerland \\
and \\
Instituut-Lorentz, University of Leiden, 
P.~O.~Box 9506,\\ 
2300 RA Leiden, The Netherlands} \\
\vspace{0.5cm}
and \\
\vspace{0.5cm}
{\large W.~T.~Giele}\\
\vspace{0.5cm}
{\it
Fermi National Accelerator Laboratory, P.~O.~Box 500,\\
Batavia, IL 60510, U.S.A.} \\
\vspace{0.5cm}
\end{center}
\begin{abstract}
In this paper the general form of scattering
amplitudes for massless particles with equal
spins $s$ ($s s\rightarrow s s$) or unequal
spins ($s_a s_b \rightarrow s_a s_b$) are derived. The imposed
conditions are that the amplitudes should have 
the lowest possible dimension,
have propagators of dimension $m^{-2}$,
and obey gauge invariance. It is
shown that the number of momenta required
for amplitudes involving particles with $s>2$ is higher
than the number implied by 3-vertices for higher
spin particles derived in the literature. Therefore, 
the dimension of the coupling constants following 
from the latter 3-vertices has a smaller power
of an inverse mass than our results imply.
Consequently, the 3-vertices in the literature
cannot be the first interaction terms
of a gauge-invariant theory.
When no spins $s>2$ are present in the
process the known QCD, QED 
or (super) gravity amplitudes are obtained
from the above general amplitudes.
\end{abstract}
\begin{flushleft}
FERMILAB-Pub-97/056-T\\
CERN-TH/97-33\\
March 1997
\end{flushleft}
\vspace{1cm}
\end{titlepage}

\section{Introduction}

In the last ten years attempts have been made
to prove or disprove the existence of field
theories involving massless particles with
spin higher than 2. Free field theories for
spin $s$ particles exist \cite{r1,r2} and
the fields obey a gauge invariance which is
a generalization of spin 1 and spin 2 gauge
invariances. The introduction of spin 1 and
spin 2 self interactions is related to a 
deformation of the original algebra of gauge
transformations. Constructing a Lagrangian
for massless spin 1 or spin 2 particles can 
be done by extending step by step the
Lagrangian with trilinear, quadrilinear,
\ldots\ interactions and at the same time
adding terms to the original gauge
transformation. Symbolically, the Lagrangian ${\cal L}$
and gauge transformations on the fields $\phi$ are series in the
coupling constant $g$:
\begin{equation}\label{e1.1}
{\cal L} = {\cal L}_0 + g{\cal L}_1 + g^2{\cal L}_2 +  \cdots
\end{equation}
\begin{equation}\label{e1.2}
\delta\phi = \delta_0\phi + g\delta_1\phi + g^2\delta_2\phi + \cdots\ ,
\end{equation}
where $\delta_0\phi = \delta\xi$, with $\xi$ 
a gauge parameter. At every order in $g$ there
should be gauge invariance:
\begin{equation}\label{e1.3}
\delta {\cal L} = 0\ ,
\end{equation}
or
\begin{equation}\label{e1.4}
\delta_0{\cal L}_0 = 0
\end{equation}
\begin{equation}\label{e1.5}
\delta_0{\cal L}_1 + \delta_1{\cal L}_0 = 0
\end{equation}
\begin{equation}\label{e1.6}
\delta_2{\cal L}_0 + \delta_1{\cal L}_1 + \delta_0{\cal L}_2 = 0,\ \mbox{etc.}
\end{equation}
In the literature constructions of various ${\cal L}_1$'s 
involving fields with $s>2$ have been presented. Some studies
have used the light-front gauge \cite{r3,r7} and
others \cite{r4,r5,r6} a general covariant form.
The former makes it easier to obtain
results for general $s$, the latter gives
the possibility to study the gauge
transformations in detail.

Since the simplest theories are 
evidently preferred, the construction of
trilinear interactions always aims
to find a minimal number of
derivatives on the fields. From
various studies a pattern of the
dimensionality of the coupling constants 
for $s_1$--$s_2$--$s_2$ trilinear interactions
can be established. In table \ref{t1.1}
those cases are listed which
were constructed in a covariant form \cite{r4,r5,r6}.
They turned out to be unique. Moreover, 
the spin 3 self interaction requires 
structure constants like those in spin 1 self interactions.
\begin{table}\label{t1.1}\begin{center}
\begin{tabular}{|c|c|c|c|c|c|c|c|c|} \hline
$s_1=$ & $s_2=0$ & $s_2=\frac{1}{2}$ & $s_2=1$ & $s_2=\frac{3}{2}$ 
       & $s_2=2$ & $s_2=\frac{5}{2}$ & $s_2=3$ & $s_2=\frac{7}{2}$ \\ \hline
$0$    & -1 & 0 & 1 & 2 & 3 &   &   &   \\
$1$    &  0 & 0 & 0 & 1 & 2 &   &   &   \\
$2$    &  1 & 1 & 1 & 1 & 1 & 2 &   &   \\
$3$    &  2 & 2 & 2 & 2 & 2 & 2 & 2 &   \\
$4$    &    &   &   &   &   &   &   &   \\
\hline\hline \end{tabular}
\caption[]{The inverse dimension $d$ of the coupling
constant: $[g]=m^{-d}$ for an $s_1$--$s_2$--$s_2$ interaction
${\cal L}_1$.}
\end{center}\end{table}
Results in the light-front gauge \cite{r3,r7} confirm and
extend the pattern of the table: $d=s_1-1$ 
($s_1\geq s_2$) and $d=2 s_2-s_1-1$ ($s_1\leq s_2$). Structure
constants are required for all odd spin
self interactions. It can be verified that 
the covariant results reduce to the others
when inserting the light-front gauge.

In principle the next term ${\cal L}_2$ in eq. (\ref{e1.1}) 
should be constructed. One can try to 
do this directly or one can first study
the algebra of the gauge transformations of eq. (\ref{e1.2}).
The latter approach \cite{r5} applied to spin 3 reveals
that the algebra requires additional gauge
transformations of a different type, hinting 
at the need of again including higher 
spin fields. An explicit construction of 
an interaction ${\cal L}_2$ for four spin 3 fields
was unsuccessful \cite{r8}. From these results 
a self interacting spin 3 theory with 
a coupling constant dimension $m^{-2}$
does not seem to exist. This makes it 
unlikely that self interacting 
theories of even higher spin would exist. 

So far we only mention the attempts to 
find interacting massless higher spin
theories in flat space. In the literature
it has been argued that higher spin
gravitational interactions are non-analytic
in the cosmological constant and therefore
an expansion over the flat background 
is not possible \cite{r9}. Thus the lack of success
to couple spin 2 fields to higher spin fields could be
understood. It then also seems unlikely that
higher spin ($s>2$) self interacting theories
would exist. When expanding near the anti-de Sitter
background, higher spin gravitational interactions are
well defined \cite{r9}.

The present paper tries to shed light on these
questions in a different way.
The reasoning assumes for the time being that a gauge
theory describing spin $s$ (and maybe other spins)
exists, so that a scattering amplitude can be calculated.
The scattering amplitude will originate from 3-vertices 
between two spin $s$ particles and one other particle
and a 4-vertex between four spin $s$ particles.
The propagators give all kinds of contractions between 
the momenta and polarization tensors in the vertices
and give in principle poles in the Mandelstam variables
$1/s$, $1/t$ or $1/u$.
Scattering amplitudes for equal and unequal spin
scattering will be constructed in such a way
that a minimum number of momenta are required and
that for different gauges in the polarization the
same amplitudes arise.

It then turns out that for all spins $s \leq 2$ amplitudes
are obtained which correspond to those of known theories,
with the exception of one case ($12 \rightarrow 12$).
As soon as particles participate with a spin $s>2$
the dimension of the amplitude increases, and in such
a way that the 3-vertices are required to have a
dimension $d$ higher than in table \ref{t1.1}. 
In other words, the problems encountered for the specific case
of spin 3 self interactions with a 3-vertex with three
derivatives are confirmed: for spins $s>2$ the constructed
interactions of table \ref{t1.1} do not lead to a gauge theory.

We show that scattering amplitudes involving particles with
$s>2$ exist when a suitable number of derivatives is allowed
for. This number is lower than trivial interactions involving
only gauge invariant field strengths would require. However, 
whether there exists a field theory that gives these scattering
amplitudes is an unanswered question.
The construction method automatically shows that certain
helicity amplitudes in known theories vanish at tree level,
but can exist at one-loop level. The loop integral provides
the additional momenta required for a non-vanishing amplitude.
The method can also be used to derive general decay amplitudes of a
massive particle into massless ones, e.g. 
$\pi_0\rightarrow 3\gamma$.

The outline of the paper is as follows. In section 2 the polarization
tensors for massless spin $s$ particles are described and various
gauges for polarization vectors are listed.
Section 3 is devoted to the actual construction of scattering
amplitudes.
In section 4 comparisons with some known scattering and decay
amplitudes are made. Section 5 summarizes our conclusions.

\section{Polarization states and gauge choice}

For the description of massless particles
with spin $s$ we make repeated use of
the polarization vector $e_\mu$. For a boson
with spin $s$ the decomposition 
of the polarization tensor into
$e_{\mu_1\cdots\mu_s}=e_{\mu_1}\cdots e_{\mu_s}$
will be used; for fermions this quantity will be
multiplied by a spinor in order to describe
an $s+\frac{1}{2}$ state. First, we shall
deal with bosons.

It has been shown that the scattering
amplitudes for massless particles are 
conveniently described in the
Weyl-van der Waerden spinor formalism
\cite{r10,r11}. Vectors are translated into
bispinors through
\begin{equation}\label{e2.1}
V_{\dot A B}=\sigma^\mu_{\dot A B} V_\mu\ ,
\end{equation}
where $\sigma^\mu_{\dot A B}$ are Pauli matrices.
For a null vector $V_{\dot A B}$ becomes a
product of two Weyl spinors. Thus the
momentum $K_\mu$ of a massless particle becomes
\begin{equation}\label{e2.2}
K_{\dot A B}=\sigma^\mu_{\dot A B} K_\mu =k_{\dot A} k_B\ .
\end{equation}
The polarization vector of a spin 1 outgoing massless
particle can be described by a bispinor \cite{r11} 
(for positive/negative helicity):
\begin{equation}\label{e2.3}
e^+_{\dot A B}=\frac{k_{\dot A}b_B}{\la kb\ra}
\end{equation}
\begin{equation}\label{e2.4}
e^-_{\dot A B}=\frac{b_{\dot A}k_B}{\la kb\ra^*}\ ,
\end{equation}
where $k$ is the spinor related to the momentum $K$ 
of the particle (cf. eq. (\ref{e2.2})) and $b$ is an
arbitrary spinor. The arbitrariness reflects the 
freedom of gauge choice and therefore we call $b$
a gauge spinor. 
A proper normalization would require an overall factor
$\sqrt{2}$, which we shall omit, since overall normalizations
will not be important for our arguments.
The antisymmetric spinor ``in-product''
is denoted by
\begin{equation}\label{e2.5}
\la pq\ra=p_A q_B \epsilon^{B A}=p_A q^A\ ,
\end{equation}
with the antisymmetric $2\times 2$ matrix $\epsilon$.
Complex conjugation gives spinors with dotted
indices, e.g.
\begin{equation}\label{e2.6}
\la pq\ra^*=p_{\dot A} q^{\dot A}\ .
\end{equation}

The gauge freedom in eqs. (\ref{e2.3}) and (\ref{e2.4}),
and therefore in $e_{\mu_1}\cdots e_{\mu_s}$, will play
an essential role in our discussion. For processes
involving massless particles with spin it is useful
to introduce the concept of a minimal
gauge \cite{r11}. It is that choice of gauge spinors
which minimizes the number of non-vanishing inner products
$(e_i\cdot e_j)$, where $e_i$ and $e_j$ are polarization
vectors describing the polarization states of particle
$i$ and $j$ in the process.

As an example, which will be often used in the following,
let us consider a scattering process involving four
massless particles with equal spins ($ss\rightarrow ss$)
or with pair-wise equal spins ($s_a s_a\rightarrow s_b s_b$).
In order to simplify the expressions all particles will
be considered as outgoing. In principle for the above
processes one has three different scattering amplitudes:
$A^{++++}$, $A^{+++-}$ and $A^{++--}$, where $\pm$ denotes
the helicity $\pm s$. The opposite helicity amplitudes
are obtained by complex conjugation.

For $A^{++++}$ a minimal gauge is the one where all $b_i$
are the same. Then all $e_i\cdot e_j$ will vanish. For
$A^{+++-}$ a minimal gauge is obtained when $b_1=b_2=b_3=k_4$
and $b_4=k_1$: again all $e_i\cdot e_j$ vanish.  For
$A^{++--}$ minimal gauges give just one non-vanishing
$e_i\cdot e_j$. In this case there are four minimal gauges of which
we list the main characteristics in
table \ref{t2.1} for subsequent use.
For each of the minimal gauges there
are two non-vanishing $K_m\cdot e_j$ for fixed $j$.
When imposing momentum conservation it means that there 
are two non-vanishing $K_m\cdot e_j$ for each $j$ with
opposite values.
\begin{table}\label{t2.1}\begin{center}
\begin{tabular}{|c|c|c|c|c|} \hline
Gauge choice & 1 & 2 & 3 & 4 \\ \hline
$b_{1,2}$ & $k_3$ & $k_4$ & $k_3$ & $k_4$ \\
$b_{3,4}$ & $k_1$ & $k_1$ & $k_2$ & $k_2$ \\ \hline
Non-vanishing & $2 e^+_2\cdot e^-_4 =$ & $2 e^+_2\cdot e^-_3 =$ 
              & $2 e^+_1\cdot e^-_4 =$ & $2 e^+_1\cdot e^-_3 =$ \\ 
$e^+_i\cdot e^-_j$ & ${\displaystyle \frac{\la 21\ra^*\la 34\ra}{\la 23\ra\la 41\ra^*}}$ 
                   & ${\displaystyle \frac{\la 21\ra^*\la 43\ra}{\la 24\ra\la 31\ra^*}}$
                   & ${\displaystyle \frac{\la 12\ra^*\la 34\ra}{\la 13\ra\la 42\ra^*}}$ 
                   & ${\displaystyle \frac{\la 12\ra^*\la 43\ra}{\la 14\ra\la 32\ra^*}}$\\ 
& & & & \\ \hline & & & & \\
$2 K_2\cdot e_1^+$ & ${\displaystyle \frac{\la 21\ra^*\la 23\ra}{\la 13\ra}}$ 
                   & ${\displaystyle \frac{\la 21\ra^*\la 24\ra}{\la 14\ra}}$ 
                   & ${\displaystyle \frac{\la 21\ra^*\la 23\ra}{\la 13\ra}}$ 
                   & ${\displaystyle \frac{\la 21\ra^*\la 24\ra}{\la 14\ra}}$\\
& & & & \\ \hline & & & & \\
$2 K_1\cdot e_2^+$ & ${\displaystyle \frac{\la 12\ra^*\la 13\ra}{\la 23\ra}}$ 
                   & ${\displaystyle \frac{\la 12\ra^*\la 14\ra}{\la 24\ra}}$ 
                   & ${\displaystyle \frac{\la 12\ra^*\la 13\ra}{\la 23\ra}}$ 
                   & ${\displaystyle \frac{\la 12\ra^*\la 14\ra}{\la 24\ra}}$\\
& & & & \\ \hline & & & & \\
$2 K_4\cdot e_3^-$ & ${\displaystyle \frac{\la 43\ra\la 41\ra^*}{\la 31\ra^*}}$ 
                   & ${\displaystyle \frac{\la 43\ra\la 41\ra^*}{\la 31\ra^*}}$ 
                   & ${\displaystyle \frac{\la 43\ra\la 42\ra^*}{\la 32\ra^*}}$ 
                   & ${\displaystyle \frac{\la 43\ra\la 42\ra^*}{\la 32\ra^*}}$\\
& & & & \\ \hline & & & & \\
$2 K_3\cdot e_4^-$ & ${\displaystyle \frac{\la 34\ra\la 31\ra^*}{\la 41\ra^*}}$ 
                   & ${\displaystyle \frac{\la 34\ra\la 31\ra^*}{\la 41\ra^*}}$ 
                   & ${\displaystyle \frac{\la 34\ra\la 32\ra^*}{\la 42\ra^*}}$ 
                   & ${\displaystyle \frac{\la 34\ra\la 32\ra^*}{\la 42\ra^*}}$\\
& & & & \\ \hline
Non-vanishing & $\la 2|e_1^+|4\ra=$ & $\la 2|e_1^+|3\ra=$ & $\la 2|e_1^+|4\ra=$ & $\la 2|e_1^+|3\ra=$ \\
 $\la m|e_i|n\ra$  & ${\displaystyle \frac{\la 21\ra^*\la 43\ra}{\la 13\ra}}$ 
              & ${\displaystyle \frac{\la 21\ra^*\la 34\ra}{\la 14\ra}}$ 
              & ${\displaystyle \frac{\la 21\ra^*\la 43\ra}{\la 13\ra}}$ 
              & ${\displaystyle \frac{\la 21\ra^*\la 34\ra}{\la 14\ra}}$ \\
& & & & \\ \cline{2-5}
& $\la 1|e_2^+|4\ra=$ & $\la 1|e_2^+|3\ra=$ & $\la 1|e_2^+|4\ra=$ & $\la 1|e_2^+|3\ra=$ \\
              & ${\displaystyle \frac{\la 12\ra^*\la 43\ra}{\la 23\ra}}$ 
              & ${\displaystyle \frac{\la 12\ra^*\la 34\ra}{\la 24\ra}}$ 
              & ${\displaystyle \frac{\la 12\ra^*\la 43\ra}{\la 23\ra}}$ 
              & ${\displaystyle \frac{\la 12\ra^*\la 34\ra}{\la 24\ra}}$ \\
& & & & \\ \cline{2-5}
& $\la 2|e_3^-|4\ra=$ & $\la 2|e_3^-|4\ra=$ & $\la 1|e_3^-|4\ra=$ & $\la 1|e_3^-|4\ra=$ \\
              & ${\displaystyle \frac{\la 21\ra^*\la 43\ra}{\la 31\ra^*}}$ 
              & ${\displaystyle \frac{\la 21\ra^*\la 43\ra}{\la 31\ra^*}}$ 
              & ${\displaystyle \frac{\la 12\ra^*\la 43\ra}{\la 32\ra^*}}$ 
              & ${\displaystyle \frac{\la 12\ra^*\la 43\ra}{\la 32\ra^*}}$ \\
& & & & \\ \cline{2-5}
& $\la 2|e_4^-|3\ra=$ & $\la 2|e_4^-|3\ra=$ & $\la 1|e_4^-|3\ra=$ & $\la 1|e_4^-|3\ra=$ \\
              & ${\displaystyle \frac{\la 21\ra^*\la 34\ra}{\la 41\ra^*}}$ 
              & ${\displaystyle \frac{\la 21\ra^*\la 34\ra}{\la 41\ra^*}}$ 
              & ${\displaystyle \frac{\la 12\ra^*\la 34\ra}{\la 42\ra^*}}$ 
              & ${\displaystyle \frac{\la 12\ra^*\la 34\ra}{\la 42\ra^*}}$ \\
& & & & \\
\hline\hline \end{tabular}
\caption[]{Minimal gauges for the amplitude $A^{++--}$}
\end{center}\end{table}

Although processes with four massless particles will be
the main focus of the paper, let us comment on minimal
gauges for a different number of particles. For two
or three particles the minimal gauges give all $e_i\cdot e_j$ 
vanishing. For processes with $n$ ($n\geq 5$) particles the
$A^{++\cdots +}$ and $A^{++\cdots +-}$ amplitudes can be
dealt with as in the $n=4$ case. The $A^{++\cdots +--}$
case has only one non-vanishing $e_i\cdot e_j$, but $n-2$
non-vanishing $K_m\cdot e_j$ for fixed $j$.

For massless particles with spin $s+\frac{1}{2}$ the
polarization multispinors are a product of the bispinors
of eqs. (\ref{e2.3}) and (\ref{e2.4}) and of a single spinor.
For outgoing particles with helicity $\pm (s+\frac{1}{2})$
and for outgoing antiparticles with helicity $\pm (s+\frac{1}{2})$
we have respectively
\begin{equation}\label{e2.4a}
(e^+)^s k_{\dot C} \mbox{, particle, } (s+\frac{1}{2})
\end{equation}
\begin{equation}\label{e2.5a}
(e^-)^s k^C \mbox{, particle, } -(s+\frac{1}{2})
\end{equation}
\begin{equation}\label{e2.6a}
(e^+)^s k^{\dot C} \mbox{, antiparticle, } (s+\frac{1}{2})
\end{equation}
\begin{equation}\label{e2.7a}
(e^-)^s k_C \mbox{, antiparticle, } -(s+\frac{1}{2})\ .
\end{equation}
The spinors of the fermions can be combined into spinorial
products amongst themselves or with vectors, e.g.
\begin{equation}\label{e2.8}
\la 1|V|2\ra \equiv k_{1\dot A} V^{\dot A B} k_{2 B} = \la 1v\ra^*\la 2v\ra\ ,
\end{equation}
when $V$ is a null vector. In table \ref{t2.1} the non-vanishing
$\la m|e_i|n\ra$ are also listed for a case with two positive and two
negative helicity particles outgoing, giving the spinorial factor
$k_{1\dot A} k_{2\dot B} k_{3 C} k_{4 D}$.

\section{The structure of massless spin scattering amplitudes}

In this section we shall consider scattering 
of massless particles with spin, where all
spins are the same or are pair-wise the
same. We shall construct those 
scattering amplitudes where a minimal
number of momenta is involved and 
which are still obeying gauge invariance. This
number of momenta will be determined by the 
spins of the particles. Once the spins
are chosen, the number of polarization
vectors is fixed. Since in a 
minimal gauge many $e_i\cdot e_j$ vanish, those
$e_{i,j}$ have to be contracted with external
momenta, of which a minimum number will be required.
Moreover, the amplitudes must be constructed 
in such a way that the different minimal gauges give
the same amplitude. 
Initially only bosons will be considered and later
on also fermions. For the bosons, let us 
start with $ss\rightarrow ss$ scattering, where
for convenience all particles are again
assumed to be outgoing.

From the discussion in section 2 
it is clear that the amplitudes
$A^{++++}$ and $A^{+++-}$ will require
$4s$ momenta factors, denoted
symbolically as $(K)^{4s}$ 
in order not to be identically zero. 
A lower number
is required for non-vanishing $A^{++--}$
amplitudes, namely $(K)^{2s}$. Depending on which
of the minimum gauges is chosen we obtain
the general expressions
\begin{eqnarray}
A^{++--} &=& c_1 \left(e_2^+\cdot e_4^-\right)^s \left(K_2\cdot e_1^+\right)^s
                 \left(K_4\cdot e_3^-\right)^s  \nonumber \\
         &=& c_2 \left(e_2^+\cdot e_3^-\right)^s \left(K_2\cdot e_1^+\right)^s
                 \left(K_3\cdot e_4^-\right)^s  \nonumber \\
         &=& c_3 \left(e_1^+\cdot e_4^-\right)^s \left(K_1\cdot e_2^+\right)^s
                 \left(K_4\cdot e_3^-\right)^s  \label{e3.1} \\
         &=& c_4 \left(e_1^+\cdot e_3^-\right)^s \left(K_1\cdot e_2^+\right)^s
                 \left(K_3\cdot e_4^-\right)^s  \nonumber\ .
\end{eqnarray}
The quantities $c_i=c_i(s,t,u)$ are determined by the 
possible pole structures. At this stage of the construction
the above number of momenta will be obtained
in a theory where the 3-vertex
for spin $s$ particles has $s$ derivatives.
Note that a 4-vertex vanishes in these minimal
gauges since it should have two derivatives 
less than the product of the two 3-vertices.
In the next stage we impose gauge invariance,
implying that it should be possible to
choose the quantities $c_i$ in such a way
that all four expressions in eq. (\ref{e3.1})
give the same result. Using the explicit
expression from table 1, the four equations simplify to
\begin{equation}\label{e3.5}
A^{++--}={\la 12\ra^{*}}^{2s}\la 34\ra^{2s}
         \left(\frac{c_1}{t^s},\frac{c_2}{u^s},\frac{c_3}{u^s},\frac{c_4}{t^s}\right)\ ,
\end{equation}
where
\begin{eqnarray}
t&=&2 K_1\cdot K_3=2 K_2\cdot K_4 \nonumber \\
u&=&2 K_1\cdot K_4=2 K_2\cdot K_3 \label{e3.6} \\
s&=&2 K_1\cdot K_2=2 K_3\cdot K_4\ . \nonumber
\end{eqnarray}

The quantities $c_i$ arise from the propagator
of the exchanged massless particles and
have the general form
\begin{equation}\label{e3.7}
c_i=\frac{\alpha_i}{s}+\frac{\beta_i}{t}+\frac{\gamma_i}{u}=\frac{P_i}{stu}\ ,
\end{equation}
where $P_i$ is a polynomial of 
second degree in $s$, $t$, $u$. Since $s+t+u=0$,
the polynomial can be written as an
arbitrary polynomial of second degree in just two Mandelstam variables,
taking the final form
\begin{equation}\label{e3.8}
P_i=A_i t^2+B_i tu+C_i u^2\ .
\end{equation}
The requirement
\begin{equation}\label{e3.10}
\frac{P_1}{t^s}=\frac{P_2}{u^s}\ \mbox{($s\geq$ 1)}
\end{equation}
can only be satisfied for
\begin{eqnarray}
s&=&1\,:\ C_1=A_2=0\,,\ A_1=B_2\,,\ B_1=C_2 \label{e3.11} \\
s&=&2\,:\ A_1=C_2\,,\ B_1=C_1=A_2=B_2=0 \label{e3.12} \ .
\end{eqnarray}
Higher $s$ values are inconsistent with eq. (\ref{e3.10}).
The amplitudes now take the form
\begin{eqnarray}
A_{s=1}^{++--} &=& {\la 12\ra^{*}}^{2}\la 34\ra^{2}
                  \left(\frac{1}{su}\,,\ \frac{1}{st}\,,\ \frac{1}{tu}\right)
                  \label{e3.13} \\
A_{s=2}^{++--} &=& \frac{{\la 12\ra^{*}}^{4}\la 34\ra^{4}}{stu}\label{e3.14} \ .
\end{eqnarray}
For spins $s > 2$ a gauge invariant theory 
with 3-vertices with $s$ derivatives does therefore
not exist. 
One may wonder what happens when the dimension of 
the propagator is $m^{-4}$ in the cases of $s=3, 4$.
One would obtain eq. (\ref{e3.7}), with the denominator squared
and a polynomial of the 4th degree with 5 coefficients
related to 3 arbitrary parameters. For the case $s=4$ 
the coefficient of $t^4$ in $P_1$ and of $u^4$ in $P_2$ should
survive, the others should vanish. This cannot be realized.
For $s=3$ eq. (\ref{e3.10}) requires at least $\beta_1=0$,
$\alpha_1+\gamma_1=0$ and $\gamma_2=0$, $\alpha_2+\beta_2=0$,
but then it is still impossible to choose $\alpha_1$ and
$\alpha_2$ such that eq. (\ref{e3.10}) is satisfied.

When the number of
derivatives is increased we can construct a 
polynomial $P_i$ of higher degree
\begin{equation}\label{e3.15}
P_i=a_{2n}^{(i)} t^{2n}+a_{2n-1}^{(i)} t^{2n-1}u+\cdots +a_{0}^{(i)} u^{2n}\ .
\end{equation}
Note that with the increased number of derivatives the
contact term between four spin $s$ particles is not
necessarily zero and therefore contributes to the
polynomial (\ref{e3.15}).
For spin $s$ the lowest degree $2n$, 
which could give a gauge invariant 
theory, will be for $n=s/2$. For even spins
we take this condition, but for odd spins we take
$n=(s+1)/2$, which generalizes eqs. (\ref{e3.13}) and (\ref{e3.14}).
This is necessary since a 3-vertex for odd spin $s$
requires an odd number of derivatives.
Gauge invariance now demands
\begin{eqnarray}
s&=&2n-1\,,\ a_{2n}^{(1)}=a_1^{(2)}\,,\ a_{2n-1}^{(1)}=a_0^{(2)}\,,\ 
                                      \mbox{all others vanish} \label{e3.16} \\
s&=&2n\,,\ a_{2n}^{(1)}=a_0^{(2)}\,,\ \mbox{all others vanish} \label{e3.17} \ .
\end{eqnarray}
The general forms will be
\begin{eqnarray}
A_{s=2n-1}^{++--} &=& {\la 12\ra^{*}}^{2s}\la 34\ra^{2s}
                  \left(\frac{1}{su}\,,\ \frac{1}{st}\,,\ \frac{1}{tu}\right)
                  \label{e3.18} \\
A_{s=2n}^{++--} &=& \frac{{\la 12\ra^{*}}^{2s}\la 34\ra^{2s}}{stu}\label{e3.19} \ .
\end{eqnarray}
From the above arguments we see that a priori a higher spin ($s>2$) 
massless gauge theory is not excluded.
The 3-vertex between odd spin $s$  
particles should have $2s-1$ derivatives ($s\geq 3$),
for even spin $2s-2$ derivatives ($s\geq 4$).
Strictly speaking, the exchanged particle is not 
necessarily a spin $s$ particle, but the 
3-vertex connecting the external spin $s$
particles with the exchanged one should
have the above number of derivatives.

At this point we would like to make three comments.
One is that we are only looking for amplitudes with 
the lowest number of derivatives. So we do not
consider ``charged'' even spin particles, which
would have $2s-1$ derivatives in 3-vertices.
The next comment is that even with this
higher number of derivatives it is still
not enough to make the $A^{++\cdots +}$ or
$A^{++\cdots +-}$ amplitudes non-vanishing.
The third comment is that for spin 1
and 2 the $n$-particle amplitude will have 
in the numerator $n-2$ and $2n-4$ momenta,
which is less than the $n$ or $2n$ needed
for non-vanishing $A^{++\cdots +}$ and $A^{++\cdots +-}$
amplitudes.

Next we consider the scattering 
process with two particles with spin $s_a$
and two with $s_b$ ($s_b > s_a > 0$) and look
for the lowest number of 
momenta required for a gauge
invariant amplitude. Again this
will come from an $A^{++--}$ amplitude
for which we now distinguish two cases
\begin{eqnarray}
A_1^{++--}&=&A^{++--}(aabb) \label{e3.20} \\
A_2^{++--}&=&A^{++--}(abab) \label{e3.21}\ .
\end{eqnarray}
The general form for $A_1$ will be for gauge choice 1
\begin{equation}\label{e3.22}
A_1^{++--}=c_1 \left(e_2^+\cdot e_4^-\right)^{s_a} \left(K_2\cdot e_1^+\right)^{s_a}
               \left(K_4\cdot e_3^-\right)^{s_b}\left(K_3\cdot e_4^-\right)^{s_b-s_a}
\end{equation}
and similar forms for the other minimal 
gauges. The lowest possible number of 
momenta is $(K)^{2s_b}$. In a similar fashion
we find, as before,
\begin{equation}\label{e3.23}
A_1^{++--}={\la 12\ra^{*}}^{2s_a}\la 34\ra^{2s_b}
         \left(\frac{c_1}{t^{s_a}},\frac{c_2}{u^{s_a}},
               \frac{c_3}{u^{s_a}},\frac{c_4}{t^{s_a}}\right)\ ,
\end{equation}
giving scattering amplitudes for $s_a=1,2$ 
that take the form
\begin{eqnarray}
{A_1}_{s_a=1}^{++--} &=& {\la 12\ra^{*}}^{2s_a}\la 34\ra^{2s_b}
                  \left(\frac{1}{su}\,,\ \frac{1}{st}\,,\ \frac{1}{tu}\right)
                  \label{e3.24} \\
{A_1}_{s_a=2}^{++--} &=& \frac{{\la 12\ra^{*}}^{2s_a}\la 34\ra^{2s_b}}{stu}\label{e3.25} \ .
\end{eqnarray}
The latter formulae generalize to odd and 
even spins $s_a$ when we increase the number of
momenta in the amplitudes (\ref{e3.22}) from $2s_b$ to
$2s_b+2(s_a-1)$ or $2s_b+2(s_a-2)$. In terms of 3-vertices
the product of two vertices has $2s_b+2(s_a-1)$ or
$2s_b+2(s_a-2)$ derivatives.

For the other amplitude $A_2^{++--}$ we start with the minimal gauge 4:
\begin{equation}\label{e3.26}
A_2^{++--}=c_4 \left(e_1^+\cdot e_3^-\right)^{s_a} \left(K_1\cdot e_2^+\right)^{s_b}
               \left(K_3\cdot e_4^-\right)^{s_b}\ .
\end{equation}
This gauge requires in the amplitude $2s_b$ momenta; the minimal
gauge 1 would require $2s_a$, but since both gauges should 
be acceptable we need at least $2s_b$ momenta. For gauges 2 and 3
we have the general form
\begin{eqnarray}\label{e3.27}
A_2^{++--}&=& c_2 \left(e_2^+\cdot e_3^-\right)^{s_a} \left(K_2\cdot e_1^+\right)^{s_a}
                 \left(K_1\cdot e_2^+\right)^{s_b-s_a}\left(K_3\cdot e_4^-\right)^{s_b} \\
          &=& c_3 \left(e_1^+\cdot e_4^-\right)^{s_a} \left(K_1\cdot e_2^+\right)^{s_b}
                 \left(K_4\cdot e_3^-\right)^{s_a}\left(K_3\cdot e_4^-\right)^{s_b-s_a}\ .
                 \label{e3.28}
\end{eqnarray}
For the given number of momenta we have in the minimal gauge 1 more
possibilities to write a general expression for $A_2^{++--}$.
The condition that $c_4$, $c_2$ and $c_3$ should be chosen in such
a way that (\ref{e3.26})--(\ref{e3.28}) give the same results fixes
the amplitude. The remaining amplitude with $c_1$ can also be
constructed. The resulting amplitudes can be found for $s_b=1,2$
and take the form
\begin{eqnarray}
{A_2}_{s_b=1}^{++--} &=& {\la 12\ra^{*}}^{2s_b}\la 14\ra^{2(s_b-s_a)}\la 34\ra^{2s_a}
                  \left(\frac{1}{su}\,,\ \frac{1}{st}\,,\ \frac{1}{tu}\right)
                  \label{e3.29} \\
{A_2}_{s_b=2}^{++--} &=& \frac{{\la 12\ra^{*}}^{2s_b}\la 14\ra^{2(s_b-s_a)}\la 34\ra^{2s_a}}{stu}
                  \label{e3.30} \ .
\end{eqnarray}
For higher $s_b$ values we find the same expressions for odd and even $s_b$,
the number of momenta in the amplitudes (\ref{e3.26})
being $4s_b-2$ for odd $s_b$ and $4s_b-4$ for even $s_b\geq 2$.
These numbers again are the number of derivatives in the product of two
3-vertices.

For completeness one also should consider the case $s_a=0$ for the
amplitude $A^{00+-}=A^{+-}$. There is only one minimal gauge with
$b_3=k_4$, $b_4=k_3$. Thus the amplitude will require at least
$(K)^{2s_b}$ momenta:
\begin{equation}\label{e3.31}
A^{+-} = c \left(K_1\cdot e_3^+\right)^{s_b}
           \left(K_1\cdot e_4^-\right)^{s_b}
       = c \frac{\left[\la 13\ra^{*}\la 14\ra\right]^{2s_b}}{s^{s_b}}\ ,
\end{equation}
where $c$ contains the propagator poles.
In another gauge it should be possible to obtain an expression
compatible with eq. (\ref{e3.31}).
Take for instance $b_{3,4}=k_1$, with
\begin{equation}\label{e3.32}
e_3^+\cdot e_4^- = \frac{\la 31\ra^*\la 14\ra}{\la 31\ra\la 41\ra^*} 
                 = -\frac{\left[\la 31\ra^*\la 14\ra\right]^2}{tu}\ .
\end{equation} 
The amplitude in this gauge is of the form
\begin{equation}\label{e3.33}
A^{+-} = d \frac{\left[\la 13\ra^{*}\la 14\ra\right]^{2s_b}}{t^{s_b}u^{s_b}} (K)^{2s_b}\ ,
\end{equation}
where the polynomial $(K)^{2s_b}$ is still arbitrary and $d$ contains the
poles. When we take for $(K)^{2s_b}$ the 
specific form $u^{s_b}$, compatibility
between (\ref{e3.31}) and (\ref{e3.33}) can be achieved for $s_b=1$ or $s_b=2$,
leading to
\begin{eqnarray}
A_{s_b=1}^{+-} &=& \left[\la 13\ra^{*}\la 14\ra\right]^{2s_b}
                  \left(\frac{1}{su}\,,\ \frac{1}{st}\,,\ \frac{1}{tu}\right)
                  \label{e3.34} \\
A_{s_b=2}^{+-} &=& \frac{\left[\la 13\ra^{*}\la 14\ra\right]^{2s_b}}{stu}
                  \label{e3.35} \ .
\end{eqnarray}
These forms generalize to higher spin when more momenta are allowed for.
The amplitude (\ref{e3.30}) requires not $2s_b$ momenta, 
but $4s_b-2$ momenta for odd
$s_b$ or $4s_b-4$ momenta for even $s_b$. Thus (\ref{e3.34}) and (\ref{e3.35})
reduce to $A_2^{0+0-}$ of eqs. (\ref{e3.29}) and (\ref{e3.30}).

Next we consider fermion--fermion scattering for equal and
unequal spins and finally fermion--boson scattering.
For fermion--fermion scattering with equal $s+1/2$ spins
we have at our disposal for the construction of
the amplitudes not only $4s$ polarization vectors
but also 4 spinors. For the construction of
amplitudes, the occurrence of these spinors could in
principle reduce the number of required momenta,
since spinorial forms can arise:
\begin{equation}\label{e3.36}
k_{\dot A}e^{\dot A B} p_B\,,\ 
k_{\dot A}e^{\dot A B}e_{\dot C B} p^{\dot C}\ ,
\end{equation}
with an odd or even number of $e$'s. We first ask whether
this possibility really reduces the number of momenta required.
Here again, we consider $A^{++++}$ and $A^{+++-}$. In the first case
we need even strings of polarization bispinors, which will be zero when all
$b_i$ are the same. For the second case we need one odd string and/or
an even one. They cannot be made non-vanishing. For these amplitudes one
still needs $4s$ momenta for a non-vanishing result.

For $A^{++--}$ the spinors are
${k_1}_{\dot A}{k_2}^{\dot B}{k_3}^{C}{k_4}_{D}$ or
${k_1}_{\dot A}{k_2}_{\dot B}{k_3}_{C}{k_4}_{D}$
for particles 2, 4 or 3, 4 being antiparticles. The question is 
whether for a minimal gauge the above spinors can reduce the number of 
momenta for a non-vanishing matrix element. Take gauge 1, so $e_1^+$
and $e_3^-$ should be contracted with the spinors. This can happen only with
${k_2}_{\dot B}{k_4}_{D}$, but then ${k_1}_{\dot A}{k_3}_{C}$
survive and must be contracted with a momentum, e.g. $K_2$. So the
number of required momenta remains the same. 
This makes the minimum number of momenta for 
$s+1/2\, s+1/2 \rightarrow s+1/2\, s+1/2$ and $s\,s\rightarrow s\,s$
scattering the same. In the latter case the amplitudes have to be
multiplied with a factor $\la 12\ra^*\la 34\ra$ 
to get the amplitudes of the former case.
Explicitly, the amplitudes could exist
for $3/2$ and $5/2$, which are obtained from
eqs. (\ref{e3.13}) and (\ref{e3.14}). For spin $s=3/2$
\begin{equation}
A_{s=3/2}^{++--} = {\la 12\ra^{*}}^{2s}\la 34\ra^{2s}
                  \left(\frac{1}{su}\,,\ \frac{1}{st}\,,\ \frac{1}{tu}\right)
                  \label{e3.37} \ .
\end{equation}
When the particles are identical, we have symmetry under $1\leftrightarrow 2$,
$3\leftrightarrow 4$ or $t\leftrightarrow u$,
and consequently
\begin{equation}\label{e3.38}
A_{s=3/2}^{++--} = \frac{{\la 12\ra^{*}}^{3}\la 34\ra^{3}}{ut}\ ,
\end{equation}
and for $s=5/2$
\begin{equation}\label{e3.39}
A_{s=5/2}^{++--} = \frac{{\la 12\ra^{*}}^{2s}\la 34\ra^{2s}}{stu}\ .
\end{equation}
These formulae generalize to higher non-integral spins,
when $s-1/2$ is odd or even.
For completeness we note that for spin $1/2$,
again taking all particles outgoing, the lowest
number of momenta is required for
\begin{equation}
A_{s=1/2}^{++--} = {\la 12\ra^{*}}\la 34\ra
                  \left(\frac{1}{s}\,,\ \frac{1}{t}\,,\ \frac{1}{u}\right)
                  \label{e3.40} \ .
\end{equation}

For unequal spin scattering for fermions
$s_a+1/2\,s_a+1/2\rightarrow s_b+1/2\,s_b+1/2$
the case of eq. (\ref{e3.20}) will still involve 
the same number of momenta. The reason is
that not all the spinors can be contracted
with $e_1, e_3, e_4$ in a non-vanishing way.
Here again, the modification for the change of 
$s_{a,b}$ into $s_{a,b}+1/2$ is a multiplication
of the amplitudes (\ref{e3.24}) and (\ref{e3.25})
by the factor $\la 12\ra^*\la 34\ra$. For the
amplitude $A_2$ the starting point (\ref{e3.26})
does not leave any room for a reduction
of the number of momenta. 
For fermions one has to multiply the
expressions (\ref{e3.29}) and (\ref{e3.30}) again
with $\la 12\ra^*\la 34\ra$. This also applies when
$s_a=0$.

In summary, for fermion--fermion scattering with spins 
$s$, $s_a$ or $s_b$ the formulae of equal and unequal
spin scattering (\ref{e3.18}), (\ref{e3.19}), (\ref{e3.24}), 
(\ref{e3.25}), (\ref{e3.29}) and (\ref{e3.30}) are valid.
The choice between odd/even cases is made depending
on $s-1/2$, $s_a-1/2$, $s_b-1/2$ being odd or even.

It is for fermion--boson scattering that the arguments
differ. We distinguish the cases
\begin{equation}\label{e3.41}
s_a+\frac{1}{2}\,,\ s_a+\frac{1}{2}\,,\ s_b\,,\ s_b\ ,
\end{equation}
\begin{equation}\label{e3.42}
s_a\,,\ s_a\,,\ s_b+\frac{1}{2}\,,\ s_b+\frac{1}{2}\ .
\end{equation}

For the $A_1^{++--}$ we get the spinorial factor
${k_1}_{\dot A}{k_2}_{\dot B}$ whereas for $A_2^{++--}$
we get ${k_1}_{\dot A}{k_3}_{B}$ (or
${k_3}_{A}{k_4}_{B}$ and ${k_2}_{\dot A}{k_4}_{B}$
for (\ref{e3.42})).
These factors should be used to reduce the number of
momenta in the expressions (\ref{e3.22}) and  (\ref{e3.26}),
respectively.
For $A_1^{++--}$, only in case (\ref{e3.42}) is this possible,
leading to
\begin{eqnarray}
{A_1}^{++--} &=& {\la 12\ra^{*}}^{2s_a}\la 34\ra^{2s_b-1}
                  \left(\frac{1}{su}\,,\ \frac{1}{st}\,,\ \frac{1}{tu}\right)
                  \label{e3.43} \\
{A_1}^{++--} &=& \frac{{\la 12\ra^{*}}^{2s_a}\la 34\ra^{2s_b-1}}{stu}
                  \label{e3.44} \ ,
\end{eqnarray}
for odd and even $s_a$. For case (\ref{e3.41}) the two
spinors give a factor $\la 12\ra^*$ to expressions 
(\ref{e3.24}) and (\ref{e3.25}).

For the amplitude $A_2^{++--}$ the starting point
is (\ref{e3.26}) where either $e_2^+$ or $e_4^-$ can
be contracted with ${k_1}_{\dot A}{k_3}_{B}$ for case
(\ref{e3.41}) or with ${k_2}_{\dot A}{k_4}_{B}$ for case
(\ref{e3.42}). Only in the former case can this be done with the
result
\begin{eqnarray}
A_2^{++--} &=& {\la 12\ra^{*}}^{2s_b}\la 14\ra^{2s_b-2s_a-1}\la 34\ra^{2s_a+1}
                  \left(\frac{1}{su}\,,\ \frac{1}{st}\,,\ \frac{1}{tu}\right)
                  \label{e3.45} \\
A_2^{++--} &=& \frac{{\la 12\ra^{*}}^{2s_b}\la 14\ra^{2s_b-2s_a-1}\la 34\ra^{2s_a+1}}{stu}
                  \label{e3.46} \ ,
\end{eqnarray}
for odd and even $s_b$. For case (\ref{e3.42}) the remaining
two spinors should be contracted with a momentum giving a factor
$\la 12\ra^*\la 14\ra$ to eqs. (\ref{e3.29}) and (\ref{e3.30}).

Summarizing, for $s_a\,s_a\rightarrow s_b\,s_b$ scattering we 
can take formulae (\ref{e3.29}), (\ref{e3.30}) and
(\ref{e3.24}), (\ref{e3.25}) with one exception. When
$s_b$ is non-integer one should take in eqs. 
(\ref{e3.24}) and (\ref{e3.25}) a factor
$\la 34\ra^{2s_b-1}$ instead of $\la 34\ra^{2s_b}$.

\section{Comparison with known results}

Summarizing, we have derived scattering amplitudes for
boson--boson, fermion--fermion and boson--fermion scattering.

For equal spins we have, both for boson--boson and fermion--fermion
scattering with all particles outgoing:
\begin{itemize}
\item $s$ or $s-1/2$ is odd:
\begin{eqnarray}
A^{++--} &=& {\la 12\ra^*}^{2s}\la 34\ra^{2s}
             \left(\frac{1}{su}\,,\ \frac{1}{st}\,,\ \frac{1}{ut}\right) \nonumber \\
&\simeq& s^{2s} \left(\frac{1}{su}\,,\ \frac{1}{st}\,,\ \frac{1}{ut}\right)\ , \label{e4.1}
\end{eqnarray}
\item $s$ or $s-1/2$ is even:
\begin{eqnarray}
A^{++--} &=& \frac{{\la 12\ra^*}^{2s}\la 34\ra^{2s}}{stu}\nonumber \\
             &\simeq& \frac{s^{2s}}{stu}\ ,\label{e4.2}
\end{eqnarray}
\end{itemize}
where the symbol $\simeq$ means an equality modulo a complex phase.
The latter formulae are given to facilitate the comparison with
the literature where the phase factors often are different.
These cases should be compared to 4-gluon, 4-graviton
and 4-gravitino amplitudes.
In the comparison we shall omit overall constants.

For the 4-gluon amplitude we use the
expressions written in terms of spinorial
products from \cite{r11}:
\begin{eqnarray}
A^{++--} &=& \la 34\ra^4\left[\frac{(a_1a_2a_3a_4)}{\la 12\ra\la 23\ra\la 34\ra\la 41\ra} 
           \right. \nonumber \\ && \left.
            +\frac{(a_1a_3a_4a_2)}{\la 13\ra\la 34\ra\la 42\ra\la 21\ra}
            +\frac{(a_1a_4a_2a_3)}{\la 14\ra\la 42\ra\la 23\ra\la 31\ra}
           \right] \nonumber \\
         &=&-{\la 12\ra^*}^2\la 34\ra^2\left[\frac{(a_1a_2a_3a_4)}{\la 12\ra\la 12\ra^*\la 14\ra\la 14\ra^*} 
            \right. \nonumber \\ && \left.
            +\frac{(a_1a_3a_4a_2)}{\la 12\ra\la 12\ra^*\la 13\ra\la 13\ra^*}
            +\frac{(a_1a_4a_2a_3)}{\la 13\ra\la 13\ra^*\la 14\ra\la 14\ra^*}
                                    \right]\ ,\label{e4.3}
\end{eqnarray}
where $(a_1a_2a_3a_4)$ denotes the trace of a product
of $SU(N)$ matrices $T^{a_i}$. The amplitude
is evidently a combination of the three expressions
of eq. (\ref{e4.1}).

The 4-graviton amplitude from ref. \cite{r12} reads $s^3/ut$,
which agrees with
(\ref{e4.2}). The 4-gravitino amplitude reads $s^3/ut$ \cite{r13},
 which for identical particles is the only 
$t\leftrightarrow u$ invariant form following
from (\ref{e4.1}).

For unequal spins we have, for both fermion--fermion and boson--boson
scattering with $0<s_a<s_b$ for $A_1^{++--}=A^{++--}(aabb)$:
\begin{itemize}
\item $s_a$ or $s_a-1/2$ odd:
\begin{eqnarray}
A_1^{++--} &=& {\la 12\ra^*}^{2s_a}\la 34\ra^{2s_b}
             \left(\frac{1}{su}\,,\ \frac{1}{st}\,,\ \frac{1}{ut}\right) \nonumber \\
&\simeq& s^{s_a+s_b} \left(\frac{1}{su}\,,\ \frac{1}{st}\,,\ \frac{1}{ut}\right)\ , 
\label{e4.4}
\end{eqnarray}
\item $s_a$ or $s_a-1/2$ even:
\begin{eqnarray}
A_1^{++--} &=& \frac{{\la 12\ra^*}^{2s_a}\la 34\ra^{2s_b}}{stu}\nonumber \\
             &\simeq& \frac{s^{s_a+s_b}}{stu}\ .\label{e4.5}
\end{eqnarray}
\end{itemize}
For the other case, $A_2^{++--} = A^{++--}(abab)$
with $0\leq s_a \leq s_b$, we have for
\begin{itemize}
\item $s_b$ or $s_b-1/2$ odd:
\begin{eqnarray}
A_2^{++--} &=& {\la 12\ra^*}^{2s_b}\la 14\ra^{2s_b-2s_a}\la 34\ra^{2s_a}
             \left(\frac{1}{su}\,,\ \frac{1}{st}\,,\ \frac{1}{ut}\right) \nonumber \\
&\simeq& 
    s^{s_a+s_b} u^{s_b-s_a}\left(\frac{1}{su}\,,\ \frac{1}{st}\,,\ \frac{1}{ut}\right)\ , 
\label{e4.6}
\end{eqnarray}
\item $s_b$ or $s_b-1/2$ even:
\begin{eqnarray}
A_2^{++--} &=& \frac{{\la 12\ra^*}^{2s_b}\la 14\ra^{2s_b-2s_a}\la 34\ra^{2s_a}}{stu}\nonumber \\
             &\simeq& \frac{s^{s_a+s_b}u^{s_b-s_a}}{stu}\ .\label{e4.7}
\end{eqnarray}
\end{itemize}

The case of spin 1--spin 2 scattering can arise from 
eqs. (\ref{e4.4}) and (\ref{e4.7})
\begin{eqnarray}
A_1^{+1\,+1\,-2\,-2} &\simeq& s^3\left(\frac{1}{su}\,,\ 
\frac{1}{st}\,,\ \frac{1}{ut}\right) 
\label{e4.8}\\
A_2^{+1\,+2\,-1\,-2} &\simeq& \frac{s^3 u}{stu} = \frac{s^2}{t}\ .\label{e4.9}
\end{eqnarray}
In quantum gravity (\ref{e4.9}) is found for 
photon--graviton scattering (\cite{r12}
(where the two formulae of eq. (21) should be interchanged). 
The amplitude (\ref{e4.8})
does not arise in quantum gravity, it vanishes.
That such an amplitude exists on general grounds 
has also been noticed in \cite{r14}.
In that paper even more non-vanishing amplitudes were found: 
however, these vanish here by the condition of gauge invariance.

Spin 0--spin 1, 2 scattering can give rise to amplitude $A_2$
\begin{eqnarray}
A_2^{0\,+1\,0\,-1} &\simeq& su\left(\frac{1}{su}\,,\ 
\frac{1}{st}\,,\ \frac{1}{ut}\right) 
\label{e4.10}\\
A_2^{0\,+2\,0\,-2} &\simeq& \frac{(su)^2}{stu} = \frac{su}{t}\ .\label{e4.11}
\end{eqnarray}
The former agrees with scalar--photon scattering and
the latter is in agreement with scalar--graviton scattering \cite{r12}. 
For spin 1/2--spin 3/2 scattering one expects from (\ref{e4.6})
\begin{equation}\label{e4.12}
A_2^{+1/2\,+3/2\,-1/2\,-3/2} \simeq s^2u
         \left(\frac{1}{su}\,,\ \frac{1}{st}\,,\ \frac{1}{ut}\right)\ ,
\end{equation}
which we could not compare with other results in the literature.

For boson--fermion scattering we have 
\begin{itemize}
\item $s_b$ non-integer, $s_a$ odd
\begin{eqnarray}
A_1^{++--} &=& {\la 12\ra^*}^{2s_a}\la 34\ra^{2s_b-2}
             \left(\frac{1}{su}\,,\ \frac{1}{st}\,,\ 
             \frac{1}{ut}\right) \nonumber \\
&\simeq& 
              s^{s_a+s_b-1}\left(\frac{1}{su}\,,\ \frac{1}{st}\,,\ 
              \frac{1}{ut}\right)\ , 
\label{e4.13}
\end{eqnarray}
\item $s_b$ non-integer, $s_a$ even
\begin{eqnarray}
A_1^{++--} &=& \frac{{\la 12\ra^*}^{2s_a}\la 34\ra^{2s_b-2}}{stu}\nonumber \\
&\simeq& \frac{s^{s_a+s_b-1}}{stu}\ ,
\label{e4.14}
\end{eqnarray}
\end{itemize}
where $s_b$ is the spin of the fermions 
and $s_a$ is odd or even.
For all other cases, eqs. (\ref{e4.4}), (\ref{e4.5}) ($s_a$ integer,
$s_b$ half-integer) and eqs. (\ref{e4.6}), (\ref{e4.7}) ($s_a$ integer,
$s_b$ half-integer or reverse) apply. Thus from eq. (\ref{e4.7}) one gets
\begin{eqnarray}
A_2^{1/2\ 2\ -1/2\ -2}&\simeq&\frac{s^{5/2}u^{3/2}}{stu}=\frac{s\sqrt{su}}{t}
\label{e4.17}\\
A_2^{3/2\ 2\ -3/2\ -2}&\simeq&\frac{s^{7/2}u^{1/2}}{stu}=\frac{s^2\sqrt{su}}{tu}
\label{e4.18}
\end{eqnarray}
in agreement with gravity (at least with the unpolarized cross section
in refs. \cite{r15,r16}) and supergravity \cite{r13}.

From (\ref{e4.6}) we have
\begin{equation}\label{e4.19}
A_2^{1/2\ 1\ -1/2\ -1}\simeq s^{3/2}u^{1/2}
    \left(\frac{1}{su}\,,\ \frac{1}{st}\,,\ \frac{1}{ut}\right)\ ,
\end{equation}
which is of a form that arises in 
quark--gluon scattering or electron--photon scattering.

So far we have compared the general scattering amplitudes
derived in section 3 to those obtained in existing gauge
theories. We would like to stress that the type of reasoning used 
in section 3 can also be useful for finding amplitudes
that arise from certain effective Lagrangians. We illustrate
this by considering an explicit example. Suppose a particular
theory makes $\pi^0\rightarrow 3\,\gamma$ decay possible. One
should find the simplest possible Lagrangian involving 3
fields $A_\mu$ and a pion field $\varphi$. From the Lagrangian
one can derive the matrix element by the method of section 3.
In an effective field theory the $A_\mu$ fields come in $F_{\mu\nu}$
combinations, i.e. $K_\mu\varepsilon_\nu-K_\nu\varepsilon_\mu$. In the 
Weyl-van der Waerden formalism these combinations become
\begin{eqnarray}
K_\mu\varepsilon_\nu^+-K_\nu\varepsilon_\mu^+ \rightarrow 
k_{\dot A} k_{\dot C}\epsilon_{BD}\label{e4.20}\ , \\
K_\mu\varepsilon_\nu^--K_\nu\varepsilon_\mu^- \rightarrow 
k_{B} k_{D}\epsilon_{\dot A\dot C}\label{e4.21}\ .
\end{eqnarray}
From these expressions it is clear that an odd number of photons
with equal helicity cannot arise. So we have to consider amplitudes
$A^{++-}$. When using a minimal gauge 1 from table \ref{t2.1} the
amplitude takes the form
\begin{eqnarray}
A^{++-} &=& c \left(K_2\cdot e_1^+\right)\left(K_1\cdot e_2^+\right)
              \left(K_2\cdot e_3^-\right)\nonumber \\
        &=& c\frac{{\la 21\ra^*}^3\la 23\ra\la 13\ra}
                  {\left(1\cdot 3\right)}\ ,\label{e4.22}
\end{eqnarray}
where $c$ contains polynomials in $K_i\cdot K_j = \left(i\cdot j\right)$
coming from derivatives on the field strengths. Since the artificial
pole (it is not present in (\ref{e4.20}), (\ref{e4.21})) should be
absent, $c$ should contain the inner product
$\left( 1\cdot 3\right)$, such that we
have in general
\begin{equation}\label{e4.23}
A^{++-}= c^\prime {\la 12\ra^*}^3\la 23\ra\la 13\ra\ .
\end{equation}
Bose symmetry requires $A^{++-}(123)=A^{++-}(213)$.
Since $\la 12\ra^3$ is odd in $1\leftrightarrow 2$ the expression
$c^\prime$ should be as well. The simplest form for $c^\prime$ 
then leads to
\begin{equation}\label{e4.24}
A^{++-}= d \left(K_1-K_2\right)\cdot K_3 {\la 12\ra^*}^3\la 23\ra\la 13\ra\ ,
\end{equation}
with an overall constant $d$ and similar expressions for
$A^{+-+}$ and $A^{-++}$.
The opposite helicity amplitudes are obtained by complex
conjugation and have the same constant $d$ when
parity is conserved. Summing over the helicities we find
\begin{eqnarray}
\sum_{\mbox{hel.}}\left|A\right|^2 &=& d^2 
\left(1\cdot 2\right)\left(2\cdot 3\right)\left(1\cdot 3\right)
\left[\left(1\cdot 2\right)^2
\left[\left(1\cdot 3\right)-\left(2\cdot 3\right)\right]^2
\right.\nonumber \\ &+& \left. 
\left(1\cdot 3\right)^2
\left[\left(1\cdot 2\right)-\left(3\cdot 2\right)\right]^2
+\left(2\cdot 3\right)^2
\left[\left(2\cdot 1\right)-\left(3\cdot 1\right)\right]^2\right]\ ,
\label{e4.25}
\end{eqnarray}
which is the same form as found from a Lagrangian \cite{r17}
\begin{equation}\label{e4.26}
{\cal L} = \partial_\alpha\varphi\varepsilon^{\mu\nu\rho\sigma}
F_{\mu\nu}\left(\partial_\gamma\partial_\beta F_{\rho\sigma}\right)
\partial^\gamma F^{\alpha\beta}
\end{equation}
\section{Discussion and Conclusions}

In order to place our results in a slightly different perspective, we 
reconsider the QCD 4-gluon amplitude. Separating the colour
structure from the momentum/helicity structures we write the
amplitude as
\begin{equation}\label{e5.1}
{\cal M}(1234) = 2ig^2\sum_{P(123)} \left(a_1a_2a_3a_4\right){\cal C}(1234)\ ,
\end{equation}
where $\left(a_1a_2a_3a_4\right)$ is a trace of a product of
$SU(N)$ matrix $T^{a_i}$. It is the construction of the gauge
invariant amplitude ${\cal C}(1234)$ in which we are interested.
It arises from three ordered Feynman diagrams and reads
\begin{eqnarray}
{\cal C}(1234)&=&2\sum_{P(123)}\frac{\left(Tr\left(F_1F_2F_3F_4\right)
                    -\frac{1}{4}Tr\left(F_1F_2\right)Tr\left(F_3F_4\right)
                    \right)}{su}
\nonumber \\ \label{e5.2}
&=&\frac{N_1}{su}\ .
\end{eqnarray}
The diagrams with a $1/u$ and $1/s$ pole have numerators with two
momenta, the constant term has none. When combining the three diagrams
there are enough momenta to allow for four field strengths. The
particular combination $N_1$ has the properties
\begin{eqnarray}
N_1(++++)&=&N_1(+++-)=0\ , \nonumber \\
N_1(++--)&=&\left[\la 12\ra^*\la 34\ra\right]^2\ .\label{e5.3}
\end{eqnarray}

For graviton--graviton scattering one has four diagrams, also a $1/t$
pole diagram. The diagrams with 3-vertices have four momenta in the
numerator, the contact term has two momenta.
Combining the four diagrams to an expression $N_2/stu$ leads to
8 momenta in $N_2$. This is sufficient for four spin 2 field strengths,
since a spin 2 field strength requires two derivatives. In momentum
space a spin 2 field strength can be written as a product of spin 1
field strengths, since $e_{\mu\nu}=e_\mu e_\nu$.
Thus a suitable form of $N_2$ would be $\left(N_1\right)^2$, being
expressible in spin 2 field strengths. Moreover, the correct
$++++$, $+++-$, $++--$ amplitudes would arise.
In fact the construction of the graviton--graviton 
scattering amplitude along these lines is explicitly
known \cite{r18,r19}:
\begin{eqnarray}
{\cal A}(1234) &=& s\times {\cal C}(1234)\, {\cal C}(1243) \nonumber \\
               &=& \frac{N_1(1234)N_1(1243)}{stu} \label{e5.4}\\
               &=& \frac{N_2}{stu}\nonumber\ .
\end{eqnarray}

Note that eq. (85) is uniquely determined by demanding
factorization of the expression in Feynman diagrams
(i.e. a term with an $s$-pole, a $u$-pole and a contact term).
So we could try to turn the reasoning around. 
Suppose we start with amplitudes like
(\ref{e5.2}) and (\ref{e5.4}). One obtains a constant term by dividing out 
the whole numerator. The $s$, $t$ or $u$ exchange diagrams are 
obtained by dividing out parts of the numerator. In this way
one gets a handle on the form of the 3- and 4-vertices.

For equal particle scattering the structure of the
amplitude we find is for odd spin
\begin{equation}\label{e5.5}
{\cal C}=\frac{N_1^s}{su}\ ,
\end{equation}
and for even spin
\begin{equation}\label{e5.6}
{\cal A}=\frac{N_2^{s/2}}{stu}\ .
\end{equation}
Since both $N_1/su$ and $N_2/stu$ can already be split into a non-pole
and pole part, also ${\cal C}$ and ${\cal A}$ can be split in such a way.
Whether one can derive acceptable 3- and 4-vertices from this and
whether the pole parts are related to only spin $s$ exchange is
to be seen.
As a side remark, one can also understand in terms of field strengths
how the difference between the $A_1$ and $A_2$ terms of eqs. (37),
(38) and (42), (43) arise. For both sets of equations one needs
a factor $N_1^{s_a}$. For the first set one then completes the
matrix element by adding a factor $\{Tr(F_3^-\cdot F_4^-)\}^{s_b-s_a}$,
for the second set one needs 
$\{K_1\cdot F_2^+\cdot F_4^-\cdot K_1\}^{s_b-s_a}$,
since a non-vanishing expression without momenta is not possible.
This can be seen from eqs. (77), (78). Moreover one thus obtains
the correct spinorial factors for the $A_1$ and $A_2$ amplitudes.

So, we have derived the form of higher spin scattering amplitudes
but do not know whether there exist a field theories that lead to them.
Let us summarize what we know about the dimensions of the amplitudes.

If the three vertices for self interaction of spin $s$ bosons as 
derived in the literature would lead to a gauge theory, the number
of momenta in the product of 3-vertices in
an $ss\rightarrow ss$ scattering amplitude would be
$2s$. For $s+1/2$ fermion--fermion scattering the same dimension
would be found when spin $s$ bosons are exchanged.
For unequal spin boson scattering, $s_as_a\rightarrow s_bs_b$,
the pattern of table \ref{t1.1} would require $2s_b$ momenta.
The same holds for fermion--boson scattering,
$s_a+1/2s_a+1/2\rightarrow s_bs_b$ and $s_as_a\rightarrow s_b+1/2s_b+1/2$.

So if the 3-vertices were to be a part of a gauge theory involving a series
of vertices with increasing number of particles, the dimension of the
scattering amplitudes would be fixed. This dimension turns out to be different
from the one found by a direct construction of scattering amplitudes
on the basis of gauge invariance and pole structure. The latter method
finds for $ss\rightarrow ss$ scattering and $s+1/2$ fermion--fermion 
scattering  that $4s-2$ or $4s-4$ momenta are required in the 
product of 3-vertices in the amplitude
for odd or even $s$. For unequal spin scattering, $s_as_a\rightarrow s_bs_b$
and $s_as_a\rightarrow s_b+1/2s_b+1/2$, $4s_b-2$ or $4s_b-4$ ($s_b$ odd/even)
momenta are needed. This latter case naturally belongs to the case of
equal spin scattering: all amplitudes $s_bs_b\rightarrow s_bs_b$ and
$s_as_a\rightarrow s_bs_b$ ($s_a<s_b$) have the same dimension.

Although the number of momenta in the 
product of 3-vertices in the amplitudes for equal spin scattering
is higher than the ones suggested by table \ref{t1.1}, it is lower
than the $4s$ momenta one would get by using just field strengths
in the vertices.
The constructed scattering amplitudes for arbitrary spins reduce
for $s\leq 2$ to
the ones of known theories with the exception of the above-mentioned 
unequal spin amplitudes with $2s_a+2s_b-2$ (4) momenta.

For the construction of the amplitudes with massless particles the
Weyl-van der Waerden formalism was again very convenient. It was
also indicated that for decay amplitudes the general form can be
easily derived within this formalism.

\end{document}